\documentclass[aps,twocolumn,pra,showpacs,floatfix]{revtex4}
\usepackage{epsfig}
\usepackage{graphicx}
\usepackage{dcolumn}
\usepackage{amssymb,amsmath}
\usepackage{mathrsfs}
\begin{document}

\title{Cm$^{15+}$ and Bk$^{16+}$ ion clocks with enhanced sensitivity to new physics}

\author{V. A. Dzuba and V. V. Flambaum}

\affiliation{School of Physics, University of New South Wales, Sydney 2052, Australia}

\begin{abstract}
We perform calculations of electronic structure of  Cm$^{15+}$ and Bk$^{16+}$ ions and demonstrate that they have transitions which combine the features of atomic optical clocks with the enhanced sensitivity to the variations of the fine structure constant $\alpha$. The high sensitivity is due to large  nuclear charge  $Z$, high ionisation degree  $Z_i$ and the effect of {\em level crossing}, which enables optical transitions between states of different configurations. These are the $6p_{1/2}-5f_{5/2} $  and $6p_{1/2}-5f_{7/2} $  transitions in the single valence electron approximation. 
Variation of $\alpha$ may be due to the interaction with scalar and pseudoscalar (axion) dark matter fields. Therefore, Cm$^{15+}$ and Bk$^{16+}$ clocks are promising  candidates to search for these fields.
\end{abstract}

\maketitle

Presently the fractional precision of the frequency measurements in optical clock transitions of Sr, Yb, Al$ ^{+} $, Hg, Hg$ ^{+} $, and Yb$ ^{+} $ atomic systems reached an unprecedented level of 10$^{-18}$ (see e.g. ~\cite{w181,w182,w183,w184,w185,w186,Ye22}). 
It is natural to use the advantage of so high accuracy of the measurements in search for the physics beyond the standard model, for example, search for the space-time variation of the fundamental constants such as the fine structure constant $\alpha$ ($\alpha = e^2/\hbar c$). Variation of $\alpha$ may be due to the interaction of electromagnetic field with scalar and pseudoscalar (axion) dark matter fields \cite{Arvanitaki,Stadnik,Stadnik2}. Therefore, the measurements of $\alpha$ variation is one of the instruments to search for dark matter. 

 The search for the slow drift  of $\alpha$ can be performed by comparing the frequencies of atomic transitions with different sensitivities to the variation of $\alpha$ over long period of time.
The  laboratory limits on the present-day time variation of $\alpha$ have already  passed  $10^{-18}$ per 
year (see e.g.  the measurements  in Refs. ~\cite{Rosenband,Leefer,Godun, Filzinger,Sherrill} and  review~\cite{Safronova1}).

The sensitivity of the atomic transitions to the variation of $\alpha$ used in these studies is not very high \cite{webbPRL,webb,Canadien}. This  is also true for all working atomic  clocks.  It was suggested in Ref.~\cite{HCI} to use optical transitions in highly-charged ions (HCI) to achieve significantly stronger sensitivity to the $\alpha$ variation. 
Indeed, effects of $\alpha$ variation in electron energies  are due to the relativistic corrections which increase proportional to $Z^2 (Z_i+1)^2$  ~\cite{webbPRL,webb,HCI}. Therefore,  to have large effects we should consider heavy highly charged ions.    To avoid cancellation  between the relativistic shifts of upper and  lower levels in an electronic transition we should consider transitions between different electronic configurations. However, frequencies of such transitions in HCI  increase proportional  to $ (Z_i+1)^2$  and are usually far  outside the laser range.  The solution is based on the fact that ordering of the electron states  depends on $Z_i$  and by removing electrons we  may achieve  so called {\em level crossing} brining the frequencies of the transitions between states of different configurations into optical region \cite{crossing}. The biggest effects happen near the crossing between $s$ and $f$, or $p_{1/2}$ and $f$ orbital energies ~\cite{webbPRL,webb,HCI,crossing}.
The search for appropriate transitions in HCI is now a popular area of theoretical \cite{hci1,Cf11,Cf1,hci4,Ho14,Cf2,Cf3,Ir17t,Os}  and experimental \cite{Bekker,Ho14e,Ir17} research   (see also  reviews~\cite{Safronova2,Sahoo}).

The highest sensitivity to $\alpha$ variation has been found in  Cf$^{17+}$ and Cf$^{15+}$ ions~\cite{Cf1,Cf11,Cf2,Cf3}.
These ions have all important factors of enhancement: high $Z$ ($Z=98$), high ionisation degree, and optical transitions between states of different configurations which  correspond to the $6p_{1/2} -5f$ transitions. The disadvantage of the use of Cf is its instability.
The most long-lived isotopes of Cf are $^{249}$Cf (the half-life is 351 years) and $^{251}$Cf (the half-life is 898 years)~\cite{isotopes}.  Isotope  $^{250}$Cf with zero nuclear spin lives only 13 years. 

In present work we study the Cm$^{15+}$ and Bk$^{16+}$ ions. They are very similar to the Cf$^{17+}$ in terms of electronic spectra and the sensitivity to the variation of $\alpha$ but have much longer-living isotopes. For example, the half-life of $^{247}$Cm is 15.6 million years, $^{248}$Cm with zero nuclear spin has half-life 0.348 million years, the half-life of $^{247}$Bk is 1380 years \cite{isotopes}. In addition, the long-living  isotopes of Cm and Bk have lower value of atomic number $A$ than the long living isotopes of Cf, which probably means that it is easier to produce them. The Cm$^{15+}$, Bk$^{16+}$ and Cf$^{17+}$ ions share another important advantage - relatively simple electronic structure. They have only one electron above closed subshells and only two states above the ground state in the optical region.
This means that the high accuracy of the calculations is possible and interpretation of the experimental spectra may be relatively simple.

\begin{figure}[tb]
	\epsfig{figure=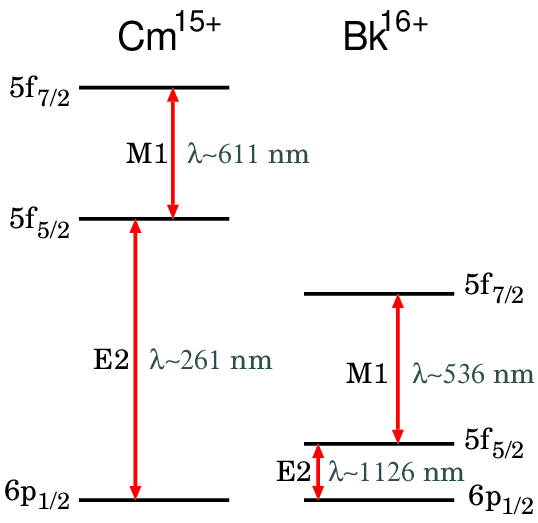,scale=0.8}
	\caption{Energy level diagram for the Cm$^{15+}$ and Bk$^{16+}$ ions (approximately in scale). 
	Possible clock transitions are shown as red double arrows.}
	\label{f:En}
\end{figure}

The energy diagram for the lowest states of Cm$^{15+}$ and Bk$^{16+}$ is presented on Fig.~\ref{f:En}.
It is based on the results of our calculations.
We perform the calculations with the use of the combination of the configuration interaction (CI) with the single-double (SD) coupled-cluster methods~\cite{SD+CIa,SD+CIb}.
The effect of various external fields (e.g., electromagnetic multipole fields for calculating transition amplitudes, hyperfine interaction, etc.) are included within the well-known random-phase approximation (RPA). The details of the calculations are presented in the Appendix. 
The results for the energy levels and  other parameters for the lowest states of Cm$^{15+}$ and Bk$^{16+}$ are presented in Table~\ref{t:E}.
The other states of the odd parity are at very high excitation energies, $E \sim $ 500000~cm$^{-1}$. 

Table~\ref{t:EE} presents the lowest energy levels of even states. Electric dipole transitions from these states to  $6s^2 6p$ and  $6s^2 5f$  may, in principle, help to measure frequency of very weak clock transitions, which can be found as the difference of frequencies of E1 transitions from one of high even state to the $6s^2 6p_{1/2}$ and $6s^2 5f_{5/2}$ states.  Moreover,  the measurement of any energy interval between the states of $6s^2 6p$ and  $6s6p5f$  configurations may help to significantly improve accuracy of the theoretical prediction of small energy interval  between the clock states.  Indeed, this interval is much smaller than the removal energy of  electron. For example,  the calculated ionisation potential of   Cm$^{15+}$  is 2,168,971 cm$^{-1}$, which is 57 times bigger than the energy interval between the $5f_{5/2}$ and $6p_{1/2}$ states. This energy interval may be treated as the difference of removal energies of $5f_{5/2}$ and $6p_{1/2}$. Therefore, one may expect that the relative theoretical error for this energy interval is  50 times bigger than the relative error in the ionisation potential. However, energy intervals within the same configuration (e.g. the fine splitting) have been calculated very accurately and reliably since there are no any cancellations here.  For example, if  the energy of optical transition  between  $6s^2 6p_{3/2}$ and any $6s6p5f$ states will be measured, we may significantly reduce  theoretical error in  the energy interval between the $5f_{5/2}$ and $6p_{1/2}$ states.

To calculate sensitivity of the atomic frequencies to the variation of $\alpha$ we present the them in the form
\begin{equation}\label{e:w}
\omega(x) = \omega_0 + q\left[\left(\frac{\alpha}{\alpha_0}\right)^2 -1\right] \equiv \omega_0 +qx,
\end{equation}
where $\omega_0$ and $\alpha_0$ are present values for the frequency and the fine structure constant, $q$ is the sensitivity coefficient.
The values of $q$ are found as a numerical derivative of the calculated values of the frequencies.
It is also convenient to have the so called enhancement factor $K$ ($K=2q/\omega$). It relates the rates of the  changing frequencies and $\alpha$. For the relative variation of the ratio of the frequencies $\omega_1/\omega_2$ we have 
\begin{equation}\label{e:q}
\frac{\dot{\omega_1}}{\omega_1} - \frac{\dot{\omega_2}}{\omega_2} = (K_1-K_2)\frac{\dot{\alpha}}{\alpha}.
\end{equation}
The calculated values of $q$ and $K$ for the lowest states of Cm$^{15+}$ and Bk$^{16+}$ ions are presented in Table~\ref{t:E}.
The values are given with respect to the ground state. These values indicate very  high sensitivity of the ionic frequencies to the variation of $\alpha$.
The values of $q$  and $K$ are comparable to  those for Cf$^{17+}$~\cite{Cf11} and much bigger than values of $q$ and $K$ in existing atomic clocks \cite{Canadien}.
 
Note however, that larger values of $q$ are more important than larger values of $K$. This is because the enhancement factor $K$ can get large value due to small value of the transition frequency (like the $6p_{1/2} - 5f_{5/2}$ transition in Bk$^{16+}$). Having small value of the transition frequency  not  always brings significant  advantage; see Ref.~\cite{Schiller} for a detailed discussion.

The lifetimes of the $5f_{5/2}$ states of the Cm$^{15+}$ and Bk$^{16+}$ ions presented in Table ~\ref{t:E} are determined by  the electric quadrupole (E2) transition to the ground state. The lifetimes of the $5f_{7/2}$ states are strongly dominated by the magnetic dipole (M1) transition to the   $5f_{5/2}$ states. See Appendix for the calculation of the amplitudes. The $5f_{5/2}$ state of Bk$^{16+}$ has very long lifetime of $\sim 800$~s due to the small frequency of the transition to the ground state (transition rate $\propto \omega^5$). The lifetimes of other states in Table~\ref{t:E}  are also relatively large. This  means that the high accuracy of the measurements is possible.

\begin{table} 
  \caption{\label{t:E} Parameters of the lowest states of Cm$^{15+}$ and Bk$^{16+}$ ions:
  excitation energies ($E$), sensitivity coefficients ($q$), enhancement factors ($K$), lifetimes ($\tau$), static dipole polarisabilities ($\alpha_0$),
  electric quadrupole moments ($Q$),  and magnetic dipole hyperfine structure constants ($A$); $g_I = \mu/I$, where $I$ is nuclear spin.}
\begin{ruledtabular}
\begin{tabular}   {l rrr cccc}
\multicolumn{1}{c}{State}&
\multicolumn{1}{c}{$E$}&
\multicolumn{1}{c}{$q$}&
\multicolumn{1}{c}{$K$}&
\multicolumn{1}{c}{$\tau$}&
\multicolumn{1}{c}{$\alpha_0$}&
\multicolumn{1}{c}{$Q$}&
\multicolumn{1}{c}{$A/g_I$}\\

&\multicolumn{1}{c}{cm$^{-1}$}
&\multicolumn{1}{c}{cm$^{-1}$}&
&&\multicolumn{1}{c}{$a_0^{3}$}&
\multicolumn{1}{c}{$|e|a_0^{2}$}&
\multicolumn{1}{c}{GHz}\\
\hline
&\multicolumn{7}{c}{Cm$^{15+}$}\\


$6p_{1/2}$ &      0 &      0 &  0 &        & 1.1998 & 0      & 154 \\    
$5f_{5/2}$ &  38375 & 367000 & 15 & 576~ms & 1.2460 & -0.459 & 1.55\\
$5f_{7/2}$ &  54730 & 380000 & 13 &  20~ms & 1.2433 & -0.550 & 0.689 \\
$6p_{3/2}$ & 182359 &        &    &        &        &        & \\

&\multicolumn{7}{c}{Bk$^{16+}$}\\

$6p_{1/2}$ &      0 &       0 &   0 &       & 1.0685 &   0    & 174 \\ 
$5f_{5/2}$ &   8880 &  403000 &  91 & 787~s & 1.1091 & -0.507 & 1.74 \\ 
$5f_{7/2}$ &  27520 &  419000 &  30 & 13~ms & 1.1069 & -0.604 & 0.772 \\
$6p_{3/2}$ & 202403 &         &     &       &        &        & \\

\end{tabular}			
\end{ruledtabular}
\end{table}

\begin{table} 
  \caption{\label{t:EE}Lowest even energy levels (cm$^{-1}$) of Cm$^{15+}$ and Bk$^{16+}$.}
\begin{ruledtabular}
\begin{tabular}   {lcr lcr}
\multicolumn{1}{c}{Conf.}&
\multicolumn{1}{c}{$J$}&
\multicolumn{1}{c}{$E$}&
\multicolumn{1}{c}{Conf.}&
\multicolumn{1}{c}{$J$}&
\multicolumn{1}{c}{$E$}\\
\hline
\multicolumn{3}{c}{Cm$^{15+}$}&
\multicolumn{3}{c}{Bk$^{16+}$}\\
$6s6p^2$ & 1/2 &   220159 &   $6s6p5f$ & 5/2 &   215399 \\
$6s6p5f$ & 5/2 &   234517 &   $6s6p5f$ & 7/2 &   228430 \\
$6s6p5f$ & 7/2 &   246936 &   $6s6p5f$ & 3/2 &   229257 \\
$6s6p5f$ & 3/2 &   250537 &   $6s6p^2$ & 1/2 &   229273 \\
$6s6p5f$ & 5/2 &   253281 &   $6s5f^2$ & 7/2 &   232137 \\
$6s6p5f$ & 7/2 &   255862 &   $6s6p5f$ & 5/2 &   233227 \\
$6s6p5f$ & 5/2 &   274931 &   $6s6p5f$ & 7/2 &   238672 \\
$6s6p5f$ & 7/2 &   275226 &   $6s5f^2$ & 3/2 &   245731 \\
$6s5f^2$ & 7/2 &   281399 &   $6s5f^2$ & 5/2 &   249308 \\
$6s5f^2$ & 3/2 &   290189 &   $6s6p5f$ & 7/2 &   256126 \\
$6s5f^2$ & 5/2 &   294383 &   $6s6p5f$ & 5/2 &   257240 \\
$6s5f^2$ & 1/2 &   314906 &   $6s5f^2$ & 1/2 &   270551 \\
$6s5f^2$ & 3/2 &   319212 &   $6s5f^2$ & 3/2 &   274819 \\
$6s5f^2$ & 3/2 &   325976 &   $6s5f^2$ & 3/2 &   282097 \\
$6s5f^2$ & 1/2 &   339563 &   $6s5f^2$ & 1/2 &   295155 \\

\end{tabular}			
\end{ruledtabular}
\end{table}

We consider now some systematic effects in the frequency measurements.

\paragraph{Blackbody radiation (BBR) shift.}
The BBR shift of the transition frequency (in Hz) is given by (see, e.g.~\cite{Derevianko})
\begin{equation}\label{e:BBR}
\delta \nu_{\rm BBR} = -8.611 \times 10^{-3}\left(\frac{T}{300K}\right)^4 \Delta \alpha_0.
\end{equation}
Here $\Delta \alpha_0$ is the difference in the static scalar polarizabilities of two states (in a.u.). We calculate the polarizabilities as described in the Appendix. The results are presented in Table~\ref{t:E}.
Using these numbers we get at T=300K the values for the relative frequency shift which are presented in Table~\ref{t:BBR}.
\begin{table} 
\caption{\label{t:BBR} Relative frequency shift due to BBR in Cm$^{15+}$ and Bk$^{16+}$}
\begin{ruledtabular}
\begin{tabular}   {ccr}
\multicolumn{1}{c}{Ion}&
\multicolumn{1}{c}{Transition}&
\multicolumn{1}{c}{$\delta \nu_{\rm BBR}/\nu$}\\
\hline
Cm$^{15+}$ & $6s_{1/2} - 5f_{5/2}$ & $-3.4\times 10^{-19}$ \\
                   & $5f_{5/2} - 5f_{7/2}$ & $ 2.2\times 10^{-20}$ \\
Bk$^{16+}$  & $6s_{1/2} - 5f_{5/2}$ & $-1.5\times 10^{-18}$ \\
                   & $5f_{5/2} - 5f_{7/2}$ & $ 3.4\times 10^{-20}$ \\
\end{tabular}			
\end{ruledtabular}
\end{table}

\paragraph{Quadrupole shift.}
The $5f$ states have sufficiently large values of the total angular momentum $J$ to make them sensitive to the gradient of electric field $\varepsilon$ via quadrupole interaction. Corresponding energy shift is given by
\begin{equation}\label{e:qu}
\Delta E_Q = \frac{J_z^2 - J(J+1)}{2J(2J-1)}Q\frac{\partial \varepsilon_z}{\partial z},
\end{equation}
where $Q$ is atomic quadrupole moment defined as doubled expectation value of the E2 operator in the stretched state
\begin{equation}\label{e:Q}
Q=2\langle J,J_z=J|E2|J,J_z=J\rangle.
\end{equation}
The calculated values of the quadrupole moment $Q$ for low states of Cm$^{15+}$ and Bk$^{16+}$ are presented in Table~\ref{t:E}.
These values are close to those calculated for Cf$^{15+}$ and Cf$^{17+}$ in Ref.~\cite{Cf3}.
Therefore, the same estimate is valid: $\delta\nu/\nu \sim 10^{-16}$. These shifts can be further suppressed by up to four orders of magnitude by averaging over projections of the total angular momentum $J$~\cite{Sr+}.

\paragraph{Hyperfine structure.} Hyperfine structure may complicate the work with the ions. In particular, it leads to enhancement of the second-order Zeeman shift, since small hyperfine intervals go into denominators of the expression for the shift. This complications can be easily avoided for Cm$^{15+}$ ion. The  isotope $^{248}$Cm lives 348 thousand years  and   has zero nuclear spin.  Bk  has no long lifetime isotopes with zero nuclear spin since it has odd number of protons. In any case, it is useful to know the values of the hyperfine constants for future analysis. We calculate magnetic dipole constant $A$ as described in the Appendix. The results are presented in Table~\ref{t:E} in a form of $A/g_I$ ($g_I = \mu/I$, $I$ is nuclear spin), which is approximately the same for all isotopes.

\paragraph{Sympathetic cooling.} Ion-based optical clocks are susceptible  to thermal motion due
to the finite ion temperature. This motion can be reduced by applying sympathetic cooling of the clock ion via the
cotrapped logic ion. The most efficient sympathetic cooling occurs when the charge-to-mass ratio of the clock ion is equal
to that of the logic ion~\cite{cooling}. The $Z_i/A$  ratio is 0.0605 for $^{248}$Cm$^{15+}$ and 0.0648 for  $^{247}$Bk$^{16+}$.
A possible logic ion in both cases is  $^{24}$Mg$^+$  where the $Z_i/A$ ratio is 0.042.

In summary, we state that the Cm$^{15+}$ and Bk$^{16+}$ ions may serve as  optical clocks which are not sensitive to external perturbations (BBR shift, quadrupole shift, etc.) but very sensitive to the hypothetical time variation of the fine structure constant and dark matter fields. 

\acknowledgments

This work was supported by the Australian Research Council Grants No. DP230101058 and DP200100150.

\appendix
\section{Method of calculations.}

We treat the Cm$^{15+}$ and Bk$^{16+}$ ions as systems with three valence electrons above the closed-shell $[1s^2 \dots 5d^{10}]$ core.
Calculations start from the relativistic Hartree-Fock (RHF) procedure for the closed-shell  core. 
The RHF Hamiltonian has the form
\begin{eqnarray}\label{e:RHF}
&&\hat H^{\rm RHF} = c \mathbf{\alpha}\cdot \mathbf{p} + (\beta-1)mc^2 + \\
&&V_{\rm nuc} + V_{\rm Breit}  + V_{\rm QED} + V_{\rm core}, \nonumber
\end{eqnarray}
where $c$ is the speed of light, $\mathbf{\alpha}$ and $\beta$ are Dirac matrixes, $\mathbf{p}$ is the electron momentum, $V_{\rm nuc}$ is the nuclear potential obtained by integrating the Fermi distribution of nuclear charge, $V_{\rm Breit}$ is the potential due to the Breit interaction~\cite{Breit}, $V_{\rm QED}$ is the potential describing  quantum electrodynamic corrections~\cite{QED}, $V_{\rm core}$ is the self-consistent RHF potential created by all core electrons.  
The single-electron basis states for valence electrons are calculated in the field of the frozen core using the B-spline technique~\cite{B-spline}. 
These basis states are used in all consequent calculations. 
This corresponds to the so-called $V^{N-M}$ approximation~\cite{VN-M}, where $M$ is the number of valence electrons ($M=3$ in our case).

The single-double coupled-cluster method (SD) is used (we use the version described in Ref.~\cite{SD+CIb}) to include the correlations between valence and core electrons. Solving the SD equations involves iterations for the core and for the valence states until the full convergence is achieved. As a result, the all-order correlation operators $\hat \Sigma_1$ and $\hat \Sigma_2$ are produced. The $\hat \Sigma_1$ operator is the single-electron operator which describes the correlation interaction of a particular valence electron with the core. The $\hat \Sigma_2$ operator is the two-electron operator which describes the screening of the Coulomb interaction between valence electrons by the core electrons. The resulting effective CI Hamiltonian has the form
\begin{equation}\label{e:HCI}
\hat H^{\rm CI} = \sum_{i=1}^M \left( \hat H^{\rm RHF} + \hat \Sigma_1\right)_i + \sum_{i<j}^M\left(\frac{e^2}{r_{ij}} + \hat \Sigma_{2ij}\right).
\end{equation}
The energy and wave function of the many-electron state $a$ are found by solving the CI equation
\begin{equation}\label{e:CI}
\left(\hat H^{\rm CI} -E_a\right)X_a=0,
\end{equation}
where $X_a$ contains coefficients of the expansion of the valence wave function over single-determinant basis states. 

To calculate transition amplitudes or hyperfine structure we need to include an additional operator of external field, such as magnetic dipole or electric quadrupole laser field, magnetic dipole nuclear field, etc. This is done in within the random-phase approximation (RPA).

The RPA equations have the form (see e.g. ~\cite{TDHF})
\begin{equation}\label{e:RPA}
(\hat H^{\rm RHF} - \epsilon_c)\delta \psi_c = - (\hat F + \delta V_{\rm core})\psi_c,
\end{equation}
where $\hat H^{\rm RHF}$ is given by (\ref{e:RHF}), index $c$ numerates single-electron states in the core, $\psi_c$ and 
$\delta \psi_c$ are corresponding single-electron functions and corrections due to the operator of external field $\hat F$, and $\delta V_{\rm core}$ is the change of the self-consistent Hartree-Fock potential due to the change in all core functions. Solving Eqs. (\ref{e:RPA}) self-consistently allows to determine  $\delta V_{\rm core}$.  Then the transition amplitude is given by  is given by
\begin{equation}\label{e:F}
A_{ab} = \langle X_b|\sum_{i=1}^M(\hat F + \delta V_{\rm core})_i |X_a\rangle,
\end{equation}
where the wave functions for states $a$ and $b$ come from solving the CI equation (\ref{e:CI}). For energy shift (e.g. in calculating the hyperfine constant) $a=b$ in (\ref{e:F}).

Polarizabilities are calculated using the wave functions found as described above and the techniques of summations over complete set of intermediate states as described in Refs~\cite{pol1,pol2}.

\end{document}